\documentclass[prc,preprint,showpacs,showkeys,nofootinbib]{revtex4}%
\usepackage{graphicx}
\usepackage{bm}
\usepackage{epsfig}
\usepackage{amsmath}
\usepackage{amsfonts}
\usepackage{amssymb}%
\begin{document}
\title{MOMDIS: a Glauber model computer code for knockout reactions}
\author{C.A. Bertulani$^{(1)}$ and A. Gade$^{(2)}$}
\email{(1) bertulani@physics.arizona.edu, (2) gade@nscl.msu.edu}
\affiliation{(1) Department of Physics, University of Arizona,
Tucson, AZ 85721, USA \\
(2) National Superconducting Cyclotron Laboratory, Michigan State
University, East Lansing, MI\ 48824, USA}
\date{\today}

\begin{abstract}
A computer program is described to calculate momentum distributions
in stripping and diffraction dissociation reactions. A Glauber model
is used with the scattering wavefunctions calculated in the eikonal
approximation. The program is appropriate for knockout reactions at
intermediate energy collisions ( 30 MeV $\leq$ E$_{lab}/$nucleon
$\leq\ 2000$ MeV). It is particularly useful for reactions involving
unstable nuclear beams, or exotic nuclei (e.g. neutron-rich nuclei),
and studies of single-particle occupancy probabilities
(spectroscopic factors) and other related physical observables. Such
studies are an essential part of the scientific program of
radioactive beam facilities, as in for instance the proposed RIA
(Rare Isotope Accelerator) facility in the US.
\footnote{\textit{``Neutron saturated nuclei are the closest one can
get to having a neutron star in the laboratory. The study of
drip-line nuclei has progressed remarkably by observing nuclear
reactions caused by radioactive fragments.'' P. Gregers Hansen,
Nature 334, 194 (1988).}\\}

\end{abstract}
\pacs{25.60.Gc, 24.50.+g, 25.60.-t }
\keywords{Direct reactions, momentum distributions, unstable nuclear beams\bigskip
\bigskip.}
\maketitle

\address{(1) Department of Physics, University of Arizona,
Tucson, AZ 85721, USA(2)\break  National Superconducting Cyclotron Laboratory, Michigan State University,
E. Lansing, MI 48824}

\narrowtext

{\Large PROGRAM SUMMARY}

\begin{enumerate}
\item \textit{Title of program: }MOMDIS (MOMentum DIStributions)

\textit{Computers:} The code has been created on an IBM-PC, but also runs on
UNIX or LINUX machines.

\textit{Operating systems:} WINDOWS or UNIX

\textit{Program language used:} Fortran-77

\textit{Memory required to execute with typical data:} 16 Mbytes of RAM memory
and 2 MB of hard disk space

\textit{No. of lines in distributed program, including test data, etc.:} 4320

\textit{Distribution format:} ASCII

\textit{Keywords:} Momentum distributions; Breakup, Stripping, Diffraction dissociation

\textit{Nature of physical problem:} The program calculates bound
wavefunctions, eikonal S-matrices, total cross sections and momentum
distributions of interest in nuclear knockout reactions at intermediate energies.

\textit{Method of solution:} Solves the radial Schr\"{o}dinger equation for
bound states. A Numerov integration is used outwardly and inwardly and a
matching at the nuclear surface is done to obtain the energy and the bound
state wavefunction with good accuracy. The S-matrices are obtained using
eikonal wavefunctions and the \textquotedblleft t-$\rho\rho$\textquotedblright%
\ method to obtain the eikonal phase-shifts. The momentum distributions are
obtained by means of a gaussian expansion of integrands. Main integrals are
performed with the Simpson's method.

\textit{Typical running time:} Almost all CPU time is consumed by calculations
of integrals, specially for transverse momentum distributions which involves
multiple integral loops. It takes up to 30 min on a 2GHz Intel P4-processor machine.
\end{enumerate}

\pagebreak

{\Large LONG WRITE-UP}

\section{Introduction}

\label{intr} Single-nucleon knockout reactions with heavy ions, at
intermediate energies and in inverse kinematics, have become a specific and
quantitative tool for studying single-particle occupancies and correlation
effects in the nuclear shell model \cite{han03}. The high sensitivity of the
method has allowed measurements on rare radioactive species available in
intensities of less than one atom per second for the incident beam. The
experiments observe reactions in which fast, mass $A$, projectiles collide
peripherally with a light nuclear target, typically $^{9}$Be or $^{12}$C,
producing residues with mass $(A-1)$, in the following referred to as the core
($c$) of the assumed two-body system of core plus nucleon. The final state of
the target and that of the struck nucleon are not observed, but instead the
energy of the final state of the residue can be identified by measuring
coincidences with decay gamma-rays emitted in flight. Referred to the
center-of-mass system of the projectile, the transferred momentum is
$\mathbf{k}_{c}$. In the sudden approximation and for the stripping reaction,
defined below, this must equal the momentum of the struck nucleon before the
collision. The measured partial cross sections to individual final levels
provide spectroscopic factors for the individual angular-momentum components
$j$. In complete analogy to the use of angular distributions in transfer
reactions, the orbital angular momentum $l$ is in the knockout reactions
revealed by the distributions of the quantity $\mathbf{k}_{c}$.

The early interest in momentum distributions came from studies of nuclear halo
states, for which the narrow momentum distributions in a qualitative way
revealed the large spatial extension of the halo wave function \cite{Ta85}. It
was shown by Bertulani and McVoy \cite{ber92} that the longitudinal component
of the momentum (taken along the beam or $z$ direction) gave the most accurate
information on the intrinsic properties of the halo and that it was
insensitive to details of the collision and the size of the target. In
contrast to this, the transverse distributions of the core are significantly
broadened by diffractive effects and by Coulomb scattering. For experiments
that observe the nucleon produced in elastic breakup, the transverse momentum
is entirely dominated by diffractive effects, as illustrated \cite{ann94} by
the angular distribution of the neutrons from the reaction $^{9}$Be($^{11}%
$Be,$^{10}$Be+n)X. In this case, the width of the transverse momentum
distribution reflects essentially the size of the target.

The cross section for the production of a given final state of the residue has
two contributions. The most important of the two, commonly referred to as
stripping or inelastic breakup, represents all events in which the removed
nucleon reacts with and excites the target from its ground state. The second
component, called diffractive or elastic breakup \cite{BB86}, represents the
dissociation of the nucleon from the residue through their two-body
interactions with the target, each being at most elastically scattered. These
events result in the removed nucleon being present in the forward beam with
essentially the beam velocity, and the target remaining in its ground state.
These processes lead to different final states, they are incoherent, and their
cross sections must be added in measurements where only the residue is
observed. General expressions for the total and differential cross sections
for the two components have been given by Hussein and McVoy \cite{HM85} and
further developed by other authors in studies of reactions with halo nuclei
(see, e.g. ref. \cite{hen96}).

In a subsequent development, the knockout method was extended to
non-halo states
\cite{nav98,tos99,tos01,nav00,aum00,mad01,end03,gad04}. For these,
involving more deeply-bound nucleons, the one-nucleon absorption
cross sections are much smaller than the free-nucleon reaction cross
section on the same target; a ratio that gives a measure of how much
the nucleon wave function is \textquotedblleft
shielded\textquotedblright\ from the target by the bulk of the core.
This required a more elaborate theoretical treatment based on the
elastic $S$-matrices $S_{c}$ and $S_{n}$ \cite{alk96,tos97} of the
core and nucleon. Other advanced theoretical treatments by
Bonaccorso and collaborators \cite{BBr88,BBo01,BBVM04} have shown
the relevance of transfer to the continuum in eikonal methods for
stripping and diffraction dissociation.

The code described in this article is based on the work presented in ref.
\cite{BH04} and on previous works, as those mentioned above.\ The code
calculates single-particle bound state wavefunctions and eikonal $S$-matrices.
Using these wavefunctions and $S$-matrices, one obtains single-particle cross
sections after integration over momentum and summation over the $m_{l}$
substates. \ As shown in ref. \cite{BH04}, the use of this code demonstrates
(i) that the transverse momentum distributions are quantitatively and even
qualitatively different from the parallel momentum distributions, and (ii) can
serve to extract angular-momentum information from the angular distributions
of the residues. The code is also of importance for calculating acceptance
corrections in experiments, and one can evaluate the correlations between
longitudinal and radial components \cite{BH04}.

\section{Bound states}

The computer code MOMDIS calculates various quantities of interest for
knockout reactions of the type%
\begin{equation}
(c+v)+T\longrightarrow c+X.\label{apb}%
\end{equation}
The internal structure of nucleus $c$, the valence particle $v$ and of the
target $T$ is not taken into account. The initial state of the projectile
nucleus $P=c+v$ is obtained by the solution of the Schr\"{o}dinger equation
for the relative motion of $c$ and $v$ in a nuclear + Coulomb potential.
Particles $c$, $v$, and $P$ have intrinsic spins labeled by $I_{c}$, $I_{v}$
and $J$, respectively. The corresponding magnetic substates are labeled by
$M_{c}$, $M_{v}$ and $M$. The orbital angular momentum for the relative motion
of $c+v$ is described by $l$ and $m$. In most situations of interest, particle
$v$ is a nucleon and $c$ is a \textquotedblleft core\textquotedblright%
\ nucleus. Thus it is convenient to couple angular momenta as $\mathbf{l+I}%
_{v}\mathbf{=j}$ and $\mathbf{j+I}_{c}\mathbf{=J}$, where $\mathbf{J}$ is the
channel spin. Below we also use the notation $\mathbf{s}$, instead of
$\mathbf{I}_{v},$ for the intrinsic spin of particle $v$.

The bound state wavefunctions of $P$ are specified by
\begin{equation}
\Psi_{JM}\left(  \mathbf{r}\right)  =R_{lj}^{J}\left(  r\right)
\mathcal{Y}_{JM}^{l}\ =\frac{u_{lj}^{J}\left(  r\right)  }{r}\mathcal{Y}%
_{JM}^{l}\ ,\label{psi_expansion}%
\end{equation}
where $r$ is the relative coordinate of $c$ and $v,$ $u_{lj}^{J}\left(
r\right)  $ is the radial wavefunction\ and $\mathcal{Y}_{JM}^{l}$ is the
spin-angle wavefunction%
\begin{equation}
\mathcal{Y}_{JM}^{l}=\sum_{m\ ,\ M_{c}}\left\langle jmI_{c}M_{c}%
|J{M}\right\rangle \left\vert jm\right\rangle \left\vert I_{c}M_{c}%
\right\rangle ,\ \ \ \ \ \ \mathrm{with}\ \ \ \ \ \left\vert jm\right\rangle
=\sum_{m_{l}\ ,\ m_{s}}Y_{lm_{l}}\left(  \widehat{\mathbf{r}}\right)
\chi_{m_{s}}\label{spinangle}%
\end{equation}
where $\chi_{m_{s}}$\ is the spinor wavefunction of particle $v$ and
$\left\langle jmI_{c}M_{c}|J{M}\right\rangle $ is a Clebsch-Gordan coefficient.

The ground-state wavefunction is normalized so that%
\begin{equation}
\int d^{3}r\ \left\vert \Psi_{JM}\left(  \mathbf{r}\right)  \right\vert
^{2}={\int\limits_{0}^{\infty}}dr\ \left\vert u_{lj}^{J}\left(  r\right)
\right\vert ^{2}=1.\label{norm}%
\end{equation}

The wavefunctions are calculated using a spin-orbit potential of the form%
\begin{equation}
V(\mathbf{r})=V_{0}(r)+V_{S}(r)\ (\mathbf{l.s})+V_{C}(r)\label{WStot}%
\end{equation}
where $V_{0}(r)$ and $V_{S}(r)$ are the central and spin-orbit interaction,
respectively, and $V_{C}(r)$ is the Coulomb potential of a uniform
distribution of charges:%
\begin{align}
V_{C}(r)  &  =\frac{Z_{c}Z_{v}e^{2}}{r}\ \ \ \mathrm{\ for}\ \ \ \ \ r>R_{C}%
\nonumber\\
&  =\frac{Z_{c}Z_{v}e^{2}}{2R_{C}}\left(  3-\frac{r^{2}}{R_{C}^{2}}\right)
\ \ \ \ \mathrm{for}\ \ \ \ r<R_{C},\label{coul_pot}%
\end{align}
where $Z_{i}$ is the charge number of nucleus $i=v,c$.

The nuclear potential is assumed to have a Woods-Saxon form plus a spin-orbit
interaction,%
\begin{align}
V_{0}(r)  &  =V_{0}\ f_{0}(r),\ \ \ \ \ \mathrm{and}\ \ \ \ \ V_{S}%
(r)=-\ V_{S0}\ \left(  \frac{\hbar}{m_{\pi}c}\right)  ^{2}\ \frac{1}{r}%
\ \frac{d}{dr}f_{S}(r)\nonumber\\
\mathrm{with}\ \ \ \ f_{i}(r)  &  =\left[  1+\exp\left(  \frac{r-R_{i}}{a_{i}%
}\right)  \right]  ^{-1}\ .\label{centsp}%
\end{align}
The spin-orbit interaction in Eq. \ref{centsp} is written in terms of the pion
Compton wavelength, $\hbar/m_{\pi}c=1.414$ fm. The parameters $V_{0}$,
$V_{S0}$, $R_{0}$, $a_{0},$ $R_{S0}$, and $a_{S0}$ are adjusted so that the
ground state energy $E_{B}$ (or the energy of an excited state) is reproduced.

The bound-state wavefunctions are calculated by solving the radial
Schr\"{o}dinger equation%
\begin{equation}
-\frac{\hbar^{2}}{2m_{ab}}\left[  \frac{d^{2}}{dr^{2}}-\frac{l\left(
l+1\right)  }{r^{2}}\right]  u_{lj}^{J}\left(  r\right)  +\left[  V_{0}\left(
r\right)  +V_{C}\left(  r\right)  +\left\langle \mathbf{s.l}\right\rangle
\ V_{S0}\left(  r\right)  \right]  u_{lj}^{J}\left(  r\right)  =E_{i}%
u_{lj}^{J}\left(  r\right) \label{bss}%
\end{equation}
where $\left\langle \mathbf{s.l}\right\rangle =\left[
j(j+1)-l(l+1)-s(s+1)\right]  /2$. This equation must satisfy the boundary
conditions $u_{lj}^{J}\left(  r=0\right)  =u_{lj}^{J}\left(  r=\infty\right)
=0$ which is only possible for discrete energies $E$ corresponding to the
bound states of the nuclear + Coulomb potential.

\section{Unitarity, Stripping and Diffraction Dissociation}

Using eikonal waves for scattering wave functions, one gets%
\begin{equation}
\Psi^{(-)\ast}(\mathbf{r})\Psi^{(+)}(\mathbf{r})=S\left(  b\right)
\exp\left(  i\mathbf{q.r}\right)  \ ,\label{intro2}%
\end{equation}
where $S\left(  b\right)  $ is the scattering matrix
\begin{equation}
S\left(  b\right)  =\exp\left[  i\chi(b)\right]  ,\ \ \ \ \ \ \ \ \text{with}%
\ \ \ \ \ \ \ \ \chi(b)=-{\frac{1}{\hbar\mathrm{v}}}\int_{-\infty}^{\infty
}dz\ U_{opt}(r)\ ,\label{intro3}%
\end{equation}
and $U_{opt}(\mathbf{r})$ is the appropriate optical potential for the
core+target and the neutron (or proton)+target scattering. In equation
(\ref{intro3}) $\chi(b)$ is the eikonal phase, and $r=\sqrt{b^{2}+z^{2}}$,
where $b$ is often interpreted as the impact parameter. This interpretation
arises from a comparison of the results obtained with eikonal wavefunctions
with those obtained in with classical particles colliding with a fixed impact
parameter $b$ \cite{BD04}. Nonetheless, the eikonal wavefunction is a quantum
scattering state and $b$ is the transverse coordinate associated to them. Thus
wave-mechanical effects, like smearing and interference, are accounted for properly.

As seen in eq. \ref{intro3}, in the eikonal\ approximation, the phase shift is
linear in the interaction. Then, according to Glauber \cite{Gl59}, the
S-matrix for $P(c+v)+T$ can be written as
\begin{equation}
\widehat{S}_{P}=\widehat{S}_{v}\widehat{S}_{c},\label{X.2}%
\end{equation}
where the $\widehat{S}$ is still an operator since it refers to a particle
inside the projectile and this depends upon parameters that have to be
averaged over by the ground state wave function of projectile, $\phi_{0}$. The
elastic breakup cross section can be easily calculated (the observed particle
is $c$)
\begin{equation}
\sigma_{el.bup}=\sum_{\mathbf{k}}\left\vert \left\langle \phi_{\mathbf{k}%
}\left\vert \widehat{S}_{b}\widehat{S}_{c}\right\vert \phi_{0}\right\rangle
\right\vert ^{2},\label{X.3}%
\end{equation}
where $\phi_{\mathbf{k}}$ is wave function that represents the continuum of
$c$ and $v$. Since
\[%
%TCIMACRO{\dint }%
%BeginExpansion
{\displaystyle\int}
%EndExpansion
\left\vert \phi_{\mathbf{k}}\right\rangle \left\langle \phi_{\mathbf{k}%
}\right\vert \linebreak d\mathbf{k}+\left\vert \phi_{0}\right\rangle
\left\langle \phi_{0}\right\vert =1,
\]
the above expression can be simplified
\begin{align}
\sigma_{el.bup} &  =%
%TCIMACRO{\dint }%
%BeginExpansion
{\displaystyle\int}
%EndExpansion
\left\langle \phi_{0}\left\vert \widehat{S}_{b}^{\ast}\widehat{S}_{c}^{\ast
}\right\vert \phi_{\mathbf{k}}\right\rangle \left\langle \phi_{\mathbf{k}%
}\left\vert \widehat{S}_{b}\widehat{S}_{c}\right\vert \phi_{0}\right\rangle
d\mathbf{k}\nonumber\\
&  =\left\langle \phi_{0}\left\vert \left\vert \widehat{S}_{b}\right\vert
^{2}\left\vert \widehat{S}_{c}\right\vert ^{2}\right\vert \phi_{0}%
\right\rangle -\left\vert \left\langle \phi_{0}\left\vert \widehat{S}%
_{b}\widehat{S}_{c}\right\vert \phi_{0}\right\rangle \right\vert
^{2}.\label{X.4}%
\end{align}

In so far as the inelastic break up is concerned, one realizes that the
detected particle must reach the detector intact and thus one must use a
survival probability to guarantee this. This survival probability is
$\left\vert \widehat{S}_{c}\right\vert ^{2}$. On the other hand the
interacting fragment $v$ is removed (stripped) and the probability for this to
happen is $\left(  1-\left\vert \widehat{S}_{v}\right\vert ^{2}\right)  $.
Identifying the transmission coefficient $T_{v}$ with $1-\left\vert
\widehat{S}_{v}\right\vert ^{2}$, we can write for the inelastic break up the
following expression \cite{HM85}%
\begin{equation}
\sigma_{in.bup}=\frac{\pi}{k^{2}}%
%TCIMACRO{\dsum }%
%BeginExpansion
{\displaystyle\sum}
%EndExpansion
\left\langle \phi_{0}\left\vert \left(  1-\widehat{T}_{c}\right)  \widehat
{T}_{v}\right\vert \phi_{0}\right\rangle .\label{X.5}%
\end{equation}

The above cross-section is also called the $v$ removal cross-section with the
notation $\sigma_{-v}$. Of course one may have the removal of the core (less
probable owing to the Coulomb barrier between the charged core and the
target)
\begin{equation}
\sigma_{-c}=\frac{\pi}{k^{2}}%
%TCIMACRO{\dsum }%
%BeginExpansion
{\displaystyle\sum}
%EndExpansion
\left\langle \phi_{0}\left\vert \left(  1-\widehat{T}_{v}\right)  \widehat
{T}_{c}\right\vert \phi_{0}\right\rangle .\label{X.6}%
\end{equation}
The sum of $\sigma_{-v}$ and $\sigma_{-c}$ gives
\begin{equation}
\sigma_{-v}+\sigma_{-c}=\frac{\pi}{k^{2}}%
%TCIMACRO{\dsum }%
%BeginExpansion
{\displaystyle\sum}
%EndExpansion
\left\langle \phi_{0}\left\vert \widehat{T}_{c}+\widehat{T}_{v}-2\widehat
{T}_{v}\widehat{T}_{c}\right\vert \phi_{0}\right\rangle .\label{X.7}%
\end{equation}
Summing the above with the \textit{total absorption} of $P,$ described \ here
by $\sigma_{-P}\equiv\sigma_{abs}=(\pi/k^{2})%
%TCIMACRO{\dsum }%
%BeginExpansion
{\displaystyle\sum}
%EndExpansion
\left\langle \phi_{0}\left\vert \widehat{T}_{v}\widehat{T}_{c}\right\vert
\phi_{0}\right\rangle $, gives
\begin{align}
\sigma_{-v}+\sigma_{-c}+\sigma_{-P}  &  =\frac{\pi}{k^{2}}%
%TCIMACRO{\dsum }%
%BeginExpansion
{\displaystyle\sum}
%EndExpansion
\left\langle \phi_{0}\left\vert \widehat{T}_{c}+\widehat{T}_{v}-\widehat
{T}_{v}\widehat{T}_{c}\right\vert \phi_{0}\right\rangle \nonumber\\
&  =\frac{\pi}{k^{2}}%
%TCIMACRO{\dsum }%
%BeginExpansion
{\displaystyle\sum}
%EndExpansion
\left\langle \phi_{0}\left\vert \left(  1-\left\vert S_{c}\right\vert
^{2}\left\vert S_{v}\right\vert ^{2}\right)  \right\vert \phi_{0}\right\rangle
=\sigma_{reaction}\,,\label{X.8}%
\end{align}
which confirms unitarity.

The expressions for $\sigma_{el.bup}$, eq. \ref{X.4}, and
$\sigma_{in.bup}$, eq. \ref{X.5}, have been used by several authors
to analyze the data on halo nuclei. It has been common to call
$\sigma_{el.bup}=\sigma_{dif}$, the \textit{diffractive break-up}
cross section and $\sigma_{in.up}=\sigma _{strip}$ the
\textit{stripping} cross section
\cite{BB86,Yab92,Bar93,Esb96,Ber98,tos99}. The cross sections also
depend on the single particle occupancy probability which can be
accounted for multiplying them by spectroscopic factors $S_{i}$
\cite{han03}.

\section{Eikonal S-matrices}

In the optical limit of the Glauber theory and the \textquotedblleft
t-$\rho\rho$\textquotedblright\ approximation (explained in detail in ref.
\cite{HRB91}), the eikonal phase is obtained from the nuclear ground state
densities and the nucleon-nucleon cross sections by the relation \cite{BD04}
(here we drop the operator notation)
\begin{equation}
S(b)=\exp\left[  i\chi(b)\right]  ,\ \ \ \ \ \ \text{with}\ \ \ \ \ \ \chi
_{N}(b)=\frac{1}{k_{NN}}\int_{0}^{\infty}dq\ q\ \rho_{p}\left(  q\right)
\rho_{t}\left(  q\right)  f_{NN}\left(  q\right)  J_{0}\left(  qb\right)
\ ,\label{eikphase}%
\end{equation}
where $\rho_{p,t}\left(  q\right)  $ is the Fourier transform of the nuclear
densities of the projectile and target, and $f_{NN}\left(  q\right)  $ is the
high-energy nucleon-nucleon scattering amplitude at forward angles, which can
be parametrized by \cite{ray79}%
\begin{equation}
f_{NN}\left(  q\right)  =\frac{k_{NN}}{4\pi}\sigma_{NN}\left(  i+\alpha
_{NN}\right)  \exp\left(  -\beta_{NN}q^{2}\right)  \ .\label{fnn}%
\end{equation}

In this equation $\sigma_{NN}$, $\alpha_{NN},$ and $\beta_{NN}$ are
parameters which fit the high-energy nucleon-nucleon scattering at
forward angles. These parameters were originally obtained for
$E_{N}\geq100$ MeV/nucleon \cite{ray79}. We use an additional set of
parameters which have been shown to reproduce the nucleon-nucleon
scattering data in the energy region $30$\ MeV$\ \leq E_{N}\leq2200$
MeV \cite{BD04}. \ The nucleon-nucleon amplitude $f_{NN}\left(
q\right)  $\ which enters equation \ref{eikphase} is of course not
the same as the one expressed in equation \ref{fnn} for the free
nucleon-nucleon scattering. Medium modifications (e.g., isospin
average, etc.) have to be taken into account. As shown in ref.
\cite{HRB91}, the largest medium effect is due to Pauli-blocking
\cite{Bert01}. The code MOMDIS allows (optional) one to include
Pauli blocking effects in the nucleon-nucleon cross section
$\sigma_{NN}$, according to refs. \cite{HRB91} and \cite{Bert01}.

In eq. (\ref{eikphase}) the quantities $\rho_{p}\left(  q\right)  $ and
$\ \rho_{t}\left(  q\right)  $ are calculated from the radial density
distributions, usually \cite{han03} taken to be of Gaussian shapes for light
nuclei, and of Fermi shapes for heavier nuclei with parameters taken from
experiment. For cases where more accuracy is needed, it is possible to take
the density distributions directly from Hartree-Fock calculations. A set of
experimental values of density parameters was published in ref. \cite{VJV87}.
These are useful, in particular in parametrizing the densities of stable nuclei.

For the Coulomb part of the optical potential the integral in eq.
(\ref{intro3}) diverges. One solves this by using $\chi=\chi_{N}+\chi_{C}$,
where $\chi_{N}$ is given by eq. (\ref{intro3}) without the Coulomb potential
and writing the Coulomb eikonal phase, $\chi_{C}$ as
\begin{equation}
\chi_{C}(b)=2\eta\ln(kb)\ ,\label{intro4}%
\end{equation}
where $\eta=Z_{p}Z_{t}e^{2}/\hbar\mathrm{v}$, $Z_{p}$ and $Z_{t}$ are the
charges of projectile and target, respectively, $v$ is their relative
velocity, $k$ their wavenumber in the center of mass system. Eq.
(\ref{intro4}) reproduces the exact Coulomb scattering amplitude when used in
the calculation of the elastic scattering with the eikonal approximation
\cite{BD04}:
\begin{equation}
f_{C}(\theta)={\frac{Z_{p}Z_{t}e^{2}}{2\mu v^{2}\ \sin^{2}(\theta/2)}}%
\ \exp\left\{  -i\eta\ \ln\left[  \sin^{2}(\theta/2)\right]  +i\pi+2i\phi
_{0}\right\} \label{fctheta}%
\end{equation}
where $\phi_{0}=arg\Gamma(1+i\eta/2)$. This is convenient for the numerical
calculations since, as shown below, the elastic scattering amplitude can be
written with the separate contribution of the Coulomb scattering amplitude
included. Then, the remaining integral (the second term on the r.h.s. of eq.
(\ref{simplif}) below) converges rapidly for the scattering at forward angles.

Although the Coulomb phase in eq. (\ref{intro4}) diverges at $b=0$, this does
not pose a real problem, since the strong absorption suppresses the scattering
at small impact parameters. It is also easy to correct this expression to
account for the finite charge distribution of the nucleus. For example,
assuming a uniform charge distribution with radius $R$ the Coulomb phase
becomes
\begin{align}
\chi_{C}(b)  &  =2\eta\ \left\{  \Theta(b-R)\ \ln(kb)+\Theta(R-b)\left[
\ln(kR)+\ln(1+\sqrt{1-b^{2}/R^{2}})\right.  \right. \nonumber\\
&  \left.  \left.  -\sqrt{1-b^{2}/R^{2}}-{\frac{1}{3}}(1-b^{2}/R^{2}%
)^{3/2}\right]  \right\}  \ ,\label{chico_0}%
\end{align}
where $\Theta$ is the step function. This expression is finite for $b=0$,
contrary to eq. \ref{intro4}. If one assumes a Gaussian distribution of charge
with radius $R$, appropriate for light nuclei, the Coulomb phase becomes
\begin{equation}
\chi_{C}(b)=2\eta\ \left[  \ln(kb)+{\frac{1}{2}}E_{1}(b^{2}/R^{2})\right]
\ ,\label{chico2}%
\end{equation}
where the error function $E_{1}$ is defined as
\begin{equation}
E_{1}(x)=\int_{x}^{\infty}{\frac{e^{-t}}{t}}\ dt\ .\label{chico3}%
\end{equation}
This phase also converges as $b\rightarrow0$. In eq. (\ref{chico_0})
$R=R_{p}+R_{t}$, while in eq. (\ref{chico2}) $R=\sqrt{R^{2}_{p}+R^{2}_{t}}$,
where $R_{p}$ and $R_{t}$ are the respective projectile and target radius. The
cost of using the expressions (\ref{chico_0}) and (\ref{chico2}) is that the
Coulomb scattering amplitude becomes more complicated than (\ref{fctheta}).
Moreover, it has been proven \cite{BD04} that the elastic and inelastic
scattering cross sections change very little by using eqs. (\ref{chico_0}) or
(\ref{chico2}), instead of eq. (\ref{intro4}).

In calculations involving stripping the final state Coulomb interaction
between the core and the target is taken into account by using the
eikonal-Coulomb phase shift of (\ref{intro4}) in the calculation of $S_{c}$.
However in the calculation of diffraction dissociation both $S_{c}$ and
$S_{n}$ are calculated using the eikonal-Coulomb phase shift of (\ref{intro4}).

\section{Elastic Cross Sections}

The calculation of elastic scattering amplitudes using eikonal wave functions,
eq. (\ref{intro2}), is very simple. They are given by \cite{BD04}%
\begin{equation}
f_{el}(\theta)=ik\ \int_{0}^{\infty}db\ b\ J_{0}(qb)\ \left\{  1-\exp\left[
i\chi(b)\right]  \right\}  \ ,\label{elast3}%
\end{equation}
where $q=2k\sin(\theta/2)$, and $\theta$ is the scattering angle. The elastic
scattering cross section is ${d\sigma_{el}/}d\Omega=\left\vert f_{el}%
(\theta)\right\vert ^{2}$. For numerical purposes, it is convenient to make
use of the analytical formula, eq. (\ref{fctheta}), for the Coulomb scattering
amplitude. Thus, if one adds and subtracts the Coulomb amplitude,
$f_{C}(\theta)$ in eq. (\ref{elast3}), one gets
\begin{equation}
f_{el}(\theta)=f_{C}(\theta)+ik\ \int_{0}^{\infty}db\ b\ J_{0}(qb)\ \exp
\left[  i\chi_{C}(b)\right]  \ \left\{  1-\exp\left[  i\chi_{N}(b)\right]
\right\}  \ .\label{simplif}%
\end{equation}

The advantage in using this formula is that the term $1-\exp\left[  i\chi
_{N}(b)\right]  $ becomes zero for impact parameters larger than the sum of
the nuclear radii (grazing impact parameter). Thus, the integral needs to be
performed only within a small range. In this formula, $\chi_{C}$ is given by
eq. (\ref{intro4}) and $f_{C}(\theta)$ is given by eq. (\ref{fctheta}), with
\begin{equation}
\phi_{0}=-\eta C+\sum_{j=0}^{\infty}\left(  {\frac{\eta}{j+1}}-\arctan
{\frac{\eta}{j+1}}\right)  \ ,\label{elast6}%
\end{equation}
and $C=0.5772156...$ is the Euler's constant.

The elastic cross section can be expressed in terms of the transverse momentum
by using the relationships $d\Omega\simeq d^{2}k_{\perp}/k^{2}$, and
$k_{\perp}\simeq q=2k\sin\left(  \theta/2\right)  $, valid for high-energy collisions.

\section{Momentum Distributions}

Following ref. \cite{HM85} the stripping reaction $(c+v)+A\longrightarrow c+X
$, where $c$ corresponds to a specified final state of the core, is given by%
\begin{equation}
\frac{d\sigma_{\mathrm{str}}}{d^{3}k_{c}}=\frac{1}{\left(  2\pi\right)  ^{3}%
}\frac{1}{2J+1}\sum_{M}\int d^{2}b_{v}\left[  1-\left\vert S_{v}\left(
b_{v}\right)  \right\vert ^{2}\right]  \left\vert \int d^{3}r\ e^{-i\mathbf{k}%
_{c}\mathbf{.r}}S_{c}\left(  b_{c}\right)  \Psi_{JM}\left(  \mathbf{r}\right)
\right\vert ^{2},\label{sknock}%
\end{equation}
and where $\mathbf{r\equiv}\left(  \mathbf{\mbox{\boldmath$\rho$}}%
,z,\phi\right)  =\mathbf{r}_{v}-\mathbf{r}_{c}$, so that
\begin{align}
b_{c}  &  =\left\vert \mathbf{\mbox{\boldmath$\rho$}}-\mathbf{b}%
_{v}\right\vert =\sqrt{\rho^{2}+b_{v}^{2}-2\rho\ b_{v}\cos\left(  \phi
-\phi_{v}\right)  }\nonumber\\
&  =\sqrt{r^{2}\sin^{2}\theta+b_{v}^{2}-2r\sin\theta\ b_{v}\cos\left(
\phi-\phi_{v}\right)  }.
\end{align}

Assuming an independence of the bound state wave functions on spin (i.e. no
spin-orbit term), equation \ref{sknock} can be recast to%
\begin{equation}
\frac{d\sigma_{\mathrm{str}}}{d^{3}k_{c}}=\frac{1}{\left(  2\pi\right)  ^{3}%
}\frac{1}{2l+1}\sum_{m}\int d^{2}b_{v}\left[  1-\left\vert S_{v}\left(
b_{v}\right)  \right\vert ^{2}\right]  \left\vert \int d^{3}r\ e^{-i\mathbf{k}%
_{c}\mathbf{.r}}S_{c}\left(  b_{c}\right)  \psi_{lm}\left(  \mathbf{r}\right)
\right\vert ^{2},\label{sknock2}%
\end{equation}
where the single-particle bound state wave functions for the subsystem $(c+v)
$ is specified by\ $\psi_{lm}\left(  \mathbf{r}\right)  =R_{l}\left(
r\right)  Y_{lm}\left(  \widehat{\mathbf{r}}\right)  $, where $R_{l}\left(
r\right)  $ is the radial wave function. The code MOMDIS uses eq.
\ref{sknock2} (and the subsequent ones below) to obtain the momentum
distributions. But it also allows one to calculate and use radial
wavefunctions with a spin-orbit potential term which is useful in some cases.

The cross sections for the longitudinal momentum distributions are obtained by
integrating eq. (\ref{sknock}) over the transverse component of $\mathbf{k}%
_{c}$, i.e. over $\mathbf{k}_{c}^{\perp},$ and using%
\begin{equation}
\int d^{2}\mathbf{k}_{c}^{\perp}\ \exp\left[  -i\mathbf{k}_{c}%
\mathbf{.(\mbox{\boldmath$\rho$}-\mbox{\boldmath$\rho$}}^{\prime})\right]
=\left(  2\pi\right)  ^{2}\delta\left(
\mathbf{\mbox{\boldmath$\rho$}-\mbox{\boldmath$\rho$}}^{\prime}\right)
\ .\label{DiracT}%
\end{equation}

One gets%
\begin{align}
\frac{d\sigma_{\mathrm{str}}}{dk_{z}}  &  =\frac{1}{\left(  2\pi\right)  ^{2}%
}\frac{1}{2l+1}\sum_{m}\int_{0}^{\infty}d^{2}b_{v}\ \left[  1-\left\vert
S_{v}\left(  b_{v}\right)  \right\vert ^{2}\right]  \ \ \int_{0}^{\infty}%
d^{2}\rho\ \left\vert S_{c}\left(  b_{c}\right)  \right\vert ^{2}\nonumber\\
&  \times\left\vert \int_{-\infty}^{\infty}dz\ \exp\left[  -ik_{z}z\right]
\psi_{lm}\left(  \mathbf{r}\right)  \right\vert ^{2}\ ,\label{strL}%
\end{align}
where $k_{z}$ represents the longitudinal component of $\mathbf{k}_{c}$.

For the transverse momentum distribution in cylindrical coordinates $k_{\bot
}=\sqrt{k_{x}^{2}+k_{y}^{2}}$, one uses in eq. (\ref{sknock})
\begin{equation}
\int_{-\infty}^{\infty}dk_{z}\ \exp\left[  -ik_{z}(z-z^{\prime})\right]
=2\pi\delta\left(  z-z^{\prime}\right)  \ ,\label{DiracZ}%
\end{equation}
and the result is
\begin{align}
\frac{d\sigma_{\mathrm{str}}}{d^{2}k_{\bot}}  &  =\frac{1}{2\pi}\frac{1}%
{2l+1}\ \int_{0}^{\infty}d^{2}b_{n}\ \left[  1-\left\vert S_{v}\left(
b_{v}\right)  \right\vert ^{2}\right] \nonumber\\
&  \times\sum_{m,\ p}\ \int_{-\infty}^{\infty}dz\ \left\vert \int d^{2}%
\rho\ \exp\left(  -i\mathbf{k}_{c}^{\perp}\mathbf{.\mbox{\boldmath$\rho$}}%
\right)  S_{c}\left(  b_{c}\right)  \psi_{lm}\left(  \mathbf{r}\right)
\right\vert ^{2}.\label{strT}%
\end{align}

The code MOMDIS also allows to calculate the transverse momentum distributions
in terms of the projection onto one of the Cartesian components of the
transverse momentum. This can be obtained directly from eq. (\ref{strT}), i.e.%
\begin{equation}
\frac{d\sigma_{\mathrm{str}}}{dk_{y}}=\int dk_{x}\ \frac{d\sigma
_{\mathrm{str}}}{d^{2}k_{\bot}}\left(  k_{x},k_{y}\right)  \ .\label{sigtx}%
\end{equation}

The total stripping cross section can be obtained by integrating either eq.
(\ref{strL}) or eq. (\ref{strT}). For example, from eq. (\ref{strL}), using
eq. (\ref{DiracZ}), one obtains%
\begin{align}
\sigma_{\mathrm{str}}  &  =\frac{2\pi}{2l+1}\int_{0}^{\infty}db_{v}%
\ b_{v}\ \left[  1-\left\vert S_{v}\left(  b_{v}\right)  \right\vert
^{2}\right]  \ \nonumber\\
&  \times\int d^{3}r\ \left\vert S_{c}\left(  \sqrt{r^{2}\sin^{2}\theta
+b_{v}^{2}-2r\sin\theta\ b_{v}\cos\phi}\right)  \right\vert ^{2}\ \sum
_{m}\left\vert \psi_{lm}\left(  \mathbf{r}\right)  \right\vert ^{2}\ .
\end{align}
Using the explicit form of the spherical harmonics%
\begin{align}
Y_{lm}\left(  \widehat{\mathbf{r}}\right)   &  =\left(  -1\right)  ^{m}%
\sqrt{\frac{2l+1}{4\pi}}\sqrt{\frac{\left(  l-m\right)  !}{\left(  l+m\right)
!}}\ P_{lm}\left(  \cos\theta\right)  \ e^{im\phi}\ \nonumber\\
&  =C_{lm}P_{lm}\left(  \cos\theta\right)  \ e^{im\phi}\label{Ylm}%
\end{align}
and
\begin{equation}
\mathbf{k}_{c}\cdot\mathbf{r}=k_{\bot}r\sin\theta\cos\left(  \phi_{k}%
-\phi\right)  +k_{\parallel}r\cos\theta\ ,
\end{equation}
part of the integral in (\ref{sknock}) is shown to be of the form \cite{BH04}
\begin{align}
\mathcal{F}_{lm}\left(  k_{\bot},k_{z},b_{v}\right)   &  =\int d^{3}%
r\ e^{-i\mathbf{k}_{c}\mathbf{.r}}S_{c}\left(  b_{c}\right)  \psi_{lm}\left(
\mathbf{r}\right) \nonumber\\
&  =C_{lm}\int dr\ r^{2}\ \sin\theta\ d\theta\ d\phi\exp\left\{  -i\left[
k_{\bot}r\sin\theta\cos\left(  \phi_{k}-\phi\right)  +k_{\parallel}r\cos
\theta\right]  \right\} \label{merd1}\\
&  \times S_{c}\left(  \sqrt{r^{2}\sin^{2}\theta+b_{v}^{2}-2r\sin\theta
\ b_{v}\cos\left(  \phi-\phi_{v}\right)  }\right)  R_{l}\left(  r\right)
P_{lm}\left(  \cos\theta\right)  \ e^{im\phi}.\nonumber
\end{align}
To simplify the calculations we can express $S_{c}\left(  b_{c}\right)  $ as
an expansion in terms of integrable functions. The S-matrices can be well
described by the expansion%
\begin{equation}
S_{c}\left(  b_{c}\right)  =\sum_{j}^{N}\alpha_{j}\ \exp\left[  -b_{c}%
^{2}/\beta_{j}^{2}\right]  \ ,\ \ \ \ \ \ \text{with}\ \ \ \beta_{j}%
=\frac{R_{L}}{j}\ .\label{scbc}%
\end{equation}
Good fits for realistic S-matrices are obtained with $N=20$, i.e. with 20
complex coefficients $\alpha_{j}$\ and $R_{L}=10-20$ fm, depending on the size
of the system. Since the real part of the S-matrices has the usual behavior of
$S_{c}\left(  b_{c}\right)  \sim0$ for $b_{c}\ll R$, and $S_{c}\left(
b_{c}\right)  \sim1$ for $b_{c}\gg R$, where $R$ is a generic nuclear size,
one of the coefficients of the expansion in eq. (\ref{scbc}) is $\alpha_{j}%
=1$, and $\beta_{j}=\infty$, which we take as the $j=0$ term in the expansion.

The expansion \ref{scbc} allows to calculate analytically many of the
integrals in eq. (\ref{sknock}), and one gets (see Appendix of ref.
\cite{BH04})%
\begin{equation}
\frac{d\sigma_{\mathrm{str}}}{k_{\bot}dk_{\bot}dk_{z}}=\ \frac{2\pi}{2l+1}%
\int_{0}^{\infty}db_{v}\ b_{v}\ \left[  1-\left\vert S_{v}\left(
b_{v}\right)  \right\vert ^{2}\right]  \ \sum_{m,\ p}C_{lm}^{2}\ \left\vert
\mathcal{A}_{lmp}\left(  k_{\bot},k_{z},b_{v}\right)  \right\vert
^{2},\label{fknock}%
\end{equation}
where%
\begin{align}
\mathcal{A}_{lmp}\left(  k_{\bot},k_{z},b_{v}\right)   &  =\sum_{j}\alpha
_{j}\ \exp\left[  -b_{v}^{2}/\beta_{j}^{2}\right] \nonumber\\
&  \times\int_{0}^{\infty}d\rho\ \rho\ J_{p}\left(  k_{\bot}\rho\right)
\exp\left[  -\rho^{2}/\beta_{j}^{2}\right]  \ I_{m-p}\left(  \frac{2\rho
b_{v}}{\beta_{j}^{2}}\right) \nonumber\\
&  \times\int_{-\infty}^{\infty}dz\ \exp\left[  -ik_{z}z\right]
\ R_{l}\left(  r\right)  \ P_{lm}\left(  \cos\theta\right)  \ .\label{almf2}%
\end{align}

The first term of the Equation (\ref{almf2}), with $\beta_{j}=\infty$ and
$\alpha_{j}=1$ can be calculated using $I_{\alpha}\left(  0\right)
=\delta_{\alpha}$.

Using the integral of eq. (\ref{DiracZ}) in eq. (\ref{fknock}) one gets for
the \textit{transverse momentum distribution}%
\begin{equation}
\frac{d\sigma_{\mathrm{str}}}{d^{2}k_{\bot}}=\ \frac{2\pi}{2l+1}\ \int
_{0}^{\infty}db_{v}\ b_{v}\ \left[  1-\left\vert S_{v}\left(  b_{v}\right)
\right\vert ^{2}\right]  \ \sum_{m,\ p}C_{lm}^{2}\ \int_{-\infty}^{\infty
}dz\ \left\vert \mathcal{D}_{lmp}\left(  k_{\bot},b_{v},z\right)  \right\vert
^{2},\label{sigtrans}%
\end{equation}
where
\begin{align}
\mathcal{D}_{lmp}\left(  k_{\bot},b_{v},z\right)   &  =\sum_{j=0}^{N}%
\alpha_{j}\ \exp\left[  -b_{v}^{2}/\beta_{j}^{2}\right] \\
&  \times\int_{0}^{\infty}d\rho\ \rho\ J_{p}\left(  k_{\bot}\rho\right)
\exp\left[  -\rho^{2}/\beta_{j}^{2}\right]  \ I_{m-p}\left(  \frac{2\rho
b_{v}}{\beta_{j}^{2}}\right)  R_{l}\left(  r\right)  \ P_{lm}\left(
\cos\theta\right)  \ .\nonumber
\end{align}

The momentum distributions from diffractive dissociation have nearly the same
shape as those from stripping \cite{BH04}. The code MOMDIS uses the equations
of this section to obtain the momentum distributions renormalized so that
their integral over momentum yields the total knockout cross section, eqs.
\ref{X.4} plus \ref{X.5}.

For cases where the magnetic quantum number $m$ differs from zero, we weight
the differential cross section with the multiplicity of 2, so that the sum
over all $m$ components gives the total cross section. The two-dimensional
momentum distribution does not depend on the azimuthal angle. It is convenient
to present it as a function of the parallel and the transverse momentum with
the definition
\begin{equation}
\frac{d^{2}\sigma_{\mathrm{str}}}{dk_{\bot}dk_{z}}=2\pi k_{\bot}\frac
{d^{3}\sigma_{\mathrm{str}}}{d^{2}k_{\bot}dk_{z}},\label{double}%
\end{equation}
which normalizes to the total cross section when the integration is extended
over the negative $k_{z}$ axis.

\section{Computer program and user's manual}

\subsection{Distribution}

The package is distributed in a tar.gz file and, under UNIX systems, can be unpacked as
follows:

gunzip {\it filename}.tar.gz

tar -xvf {\it filename}.tar

\bigskip

Unpacking the file generates two files, a README file, the Fortran source,
MOMDIS.FOR, and a directory, testdata. This directory contains sample input files
for a typical run of MOMDIS and files containing the expected output.

\bigskip

Option 1 - Eigenfunctions and energies

input file, a$\_$bound.txt - output file, bound.out

Option 2 - S matrices - Sn

input file, a$\_$sn.txt - output file, s$\_$sn.out.

Option 2 - S matrices - Sc

input file, a$\_$sc.txt - output file, s$\_$sc.out

Option 3 - Momentum distributions

input file, a$\_$bound.txt, bound.out, s$\_$sn.out, s$\_$sc.out

output file, sigma.out

\bigskip

The MOMDIS.FOR program file is in DOS format. To use on a unix system it should
first be converted.

dos2unix -ascii MOMDIS.FOR MOMDIS.f

\bigskip

It can be compiled using the command

g77 -o MOMDIS.exe MOMDIS.f

\bigskip
and run using the command

./MOMDIS.exe

\bigskip
The screen output produced from each of the sample runs mentioned above is shown in the
README file contained in the distribution.

\subsection{Fortran code}

The units used in the program are fm (femtometer) for distances and
MeV for energies. The output cross sections are given in millibarns
(mb), mb/(MeV/c), and mb/(MeV/c)$^2$.

The program is very fast, except for transverse and double-differential
momentum distributions, and does not require a complicated input. It is
divided in 3 modules:

\textbf{Module 1 -} Calculates of energy and wavefunction of bound
states. \textbf{Module 2 -} Calculates S-matrices. \textbf{Module 3
-} Calculates momentum distributions. If one chooses option Module
3, other options follow: \textbf{1 -} for longitudinal momentum
distributions, eq. \ref{strL}, \textbf{2 -} for transverse momentum
distributions, eq. \ref{sigtrans}, \textbf{3 -} for projected
transverse momentum distributions, eq. \ref{sigtx}, \textbf{4 -} for
double differential momentum distributions, eq. \ref{double}, and
\textbf{5 -} for elastic scattering cross section of the core with
the target, using eq. \ref{simplif}.

\subsection{Sample input file}

A sample input file is shown below. It is prepared to study the reaction
$^{15}$C$+\ ^{9}$Be$\ \rightarrow\ ^{14}$C$\ +X$ at 103 MeV/nucleon laboratory
energy. The rows starting with a symbol \textquotedblleft*\textquotedblright%
\ are not read as input and can be used in the input file at free will. They
are well suited to remind the user of the input procedure. For more details on
how to prepare an input file see the \textbf{readme} file located at the end
of the Fortran code. With the sample input below the code calculates the
\textit{bound-state wave function}. To obtain \textit{S-matrices}, one needs
to comment with a \textquotedblleft*\textquotedblright\ the input lines (4th
and 5th rows) for bound-state wavefunction and delete the \textquotedblleft%
*\textquotedblright\ from the input lines for S-matrices. The
comments in the following two input lines depend upon the option
parameters I$_{{\small SMAT}}$ , I$_{{\small DPROJ}},$ and
I$_{{\small DTARG}}$. For \textit{momentum distributions}, only the
first three rows of the sample file below are used.

\begin{center}%
\begin{tabular}
[c]{llllllllllllllll}\hline\hline
& {\small 6,} & {\small 15,} & {\small 4,} & {\small 9,} & {\small 103} &  &
& \ \ \ \ \  & {\small Z}$_{{\small P}},$ & {\small A}$_{{\small P}},$ &
{\small Z}$_{{\small T}},$ & {\small A}$_{{\small T}}${\small ,} &
{\small E}$_{lab}$ &  & \\
& {\small 6,} & {\small 14} &  &  &  &  &  &  & {\small A}$_{{\small CORE}},$
& {\small Z}$_{{\small CORE}}$ &  &  &  &  & \\
& {\small 1} &  &  &  &  &  &  &  & {\small S}$^{{\small 2}}${\small C} &  &
&  &  &  & \\
& {\small 1,} & {\small 0.05} &  &  &  &  &  &  & {\small n}$_{{\small 0}},$ &
{\small j}$_{{\small 0}},$ & {\small l}$_{{\small 0}}$ &  &  &  & \\
& ${\small -61.85}${\small ,} & {\small 2.67,} & {\small 0.6,} & {\small 0.} &
{\small 2.4,} & {\small 0.6,} & {\small 2.67} &  & {\small V}$_{{\small 0}},$
& {\small R}$_{{\small 0}},$ & {\small A,} & {\small V}$_{{\small S0}},$ &
{\small R}$_{{\small S0}},$ & {\small A}$_{{\small S0}},$ & {\small R}%
$_{{\small C}}$\\
& {\small 1} &  &  &  &  &  &  &  & {\small I}$_{{\small SMAT}}$ &  &  &  &  &
& \\
& {\small 2} &  &  &  &  &  &  &  & {\small I}$_{{\small POT}}$ &  &  &  &  &
& \\
{*\ } & {\small 50.,} & {\small 1.067,} & {\small 0.8,} & {\small 58.,} &
{\small 1.067,} & {\small 0.8} &  &  & {\small V}$_{{\small 0}},$ &
{\small R}$_{{\small 0}},$ & {\small D,} & {\small V}$_{{\small I}},$ &
{\small R}$_{{\small 0I}},$ & {\small D}$_{{\small I}}$ & \\
{*\ } & {\small rp.in} & {\small ip.in} &  &  &  &  &  &  & {\small FILE1} &
{\small FILE2} &  &  &  &  & \\
& {\small 0} &  &  &  &  &  &  &  & {\small I}$_{{\small PAULI}}$ &  &  &  &
&  & \\
& {\small 1,} & {\small 1} &  &  &  &  &  &  & {\small I}$_{{\small DPROJ}},$
& {\small I}$_{{\small DTARG}}$ &  &  &  &  & \\
& {\small 1.73,} & {\small 1.38,} & {\small 1.} &  &  &  &  &  &
${\small \alpha,}$ & ${\small \beta}\left(  b\right)  ,$ & ${\small \omega}$ &
&  &  & \\
{*\ } & {\small 2.,} & {\small 1.,} & {\small 1.,} & {\small 1.} &  &  &  &  &
{\small c,} & {\small a,} & {\small b,} & ${\small \omega}$ &  &  & \\
{*\ } & {\small 2.,} & {\small 1.,} & {\small 1.,} & {\small 0.} &  &  &  &  &
{\small c,} & {\small a,} & ${\small \omega,}$ & {\small \ I}$_{{\small DER}}$
&  &  & \\
{*\ } & {\small pd.in,} & {\small 0.6} &  &  &  &  &  &  & {\small FILE3,} &
${\small \alpha}_{{\small p}}$ &  &  &  &  & \\
& {\small 1.93,} & {\small 0.,} & {\small 0.} &  &  &  &  &  & ${\small \alpha
,}$ & ${\small \beta}\left(  b\right)  ,$ & ${\small \omega}$ &  &  &  & \\
{*\ } & {\small 2.,} & {\small 1.,} & {\small 1.,} & {\small 1.} &  &  &  &  &
{\small c,} & {\small a,} & $b,$ & ${\small \omega}$ &  &  & \\
{*\ } & {\small 2.,} & {\small 1.,} & {\small 1.,} & {\small 0} &  &  &  &  &
{\small c,} & {\small a,} & ${\small \omega,}$ & {\small I}$_{{\small DER}}$ &
&  & \\
{*\ } & {\small td.in,} & {\small 0.6} &  &  &  &  &  &  & {\small FILE4,} &
${\small \alpha}_{{\small p}}$ &  &  &  &  & \\\hline\hline
\end{tabular}

\end{center}

A more specific description follows. \textbf{Row 1}: Z$_{P}$, A$_{P}$ =
projectile charge and mass numbers, Z$_{T}$, A$_{T}$ = target charge and mass
numbers, E$_{lab}$ = laboratory energy per nucleon (MeV). \textbf{Row 2:
}$Z_{CORE}$, $A_{CORE}$ = projectile core ($^{14}$C) charge and mass number.
\textbf{Row 3: }$S^{2}C$ = Spectroscopic factor. \textbf{Row 4: }%
Single-particle bound state quantum numbers. $n_{0}$ = nodes of the
wave function (exclude origin), $j_{0}$ = total angular momentum,
$l_{0}$ = orbital angular momentum. \textbf{Row 5:} Parameters of
the Woods-Saxon potential used to calculate wavefunction following
eq. \ref{WStot}. $V_{0}$ = depth of central potential, $V_{S0}$ =
depth of spin-orbit potential, $R_{0}$ = radius parameter of the
central potential, $A$ = diffuseness of the central potential,
$R_{S0}$ = radius parameter of the spin-orbit potential, $A_{SO}$ =
diffuseness of the spin-orbit potential, $R_{C}$ = Coulomb radius
parameter (usually, $R_{C}=R_{0}$). \textbf{Row 6:} $I_{SMAT}=0$, or
1. If $I_{SMAT}=0$, S-matrix is for valence particle+target. If
$I_{SMAT}=1$, S-matrix is for core+target. \textbf{Row \ 7:}
$I_{POT}=0$, 1, or 2. If $I_{POT}=0$, optical potential is a
Woods-Saxon, with real and imaginary \ parts, $V+iV_{I}$. If
$I_{POT}=1$, optical potential is entered from an external, user
provided, file. If $I_{POT}=2$, the optical potential is built with
the t-$\rho\rho$ method. \textbf{Row 8: } If I$_{POT}$ = 0, enter
$V_{0}$ [$V_{I}$] = real part [imaginary] (both
%TCIMACRO{\TEXTsymbol{>} }%
%BeginExpansion
$>$
%EndExpansion
0) of Woods-Saxon potential. $R_{0}$ [$R_{0I}$] = radius parameter
($R=R_{0}(A_{P}^{1/3}+A_{T}^{1/3}$). $d$ [$d_{I}$] = diffuseness.
\textbf{Row 9: }Enter names of the input files for the real (FILE1)
and imaginary (FILE2) parts of the potential. In the input files,
the first row must give the number of the remaining rows which form
an ordered list of  $r$ (in fm) $\times\ V$ [$V_{I}$] (in MeV) in steps
of a constant value of $\Delta r$.  \textbf{Row 10:} If $I_{POT}=2$,
enter the following options: $I_{PAULI}=0$ for no Pauli correction
of nn cross sections,$I_{PAULI}=$ $1$ for Pauli correction of NN
cross sections. \textbf{Row 11:} Enter options for projectile and
target densities. $I_{DPROJ}$, $I_{DTARG}=1$,2,3,4,5. If I$_{DPROJ}$
= 1, projectile density is a three-parameter Gaussian density
\begin{equation}
G(r)=\left[  1+b\left(  \frac{r}{\alpha}\right)  ^{\omega}\right]  \exp\left(
-\frac{r^{2}}{\alpha^{2}}\right)  .\label{gas1}%
\end{equation}
If $I_{DPROJ}=2$, projectile density is a two-parameter Yukawa%
\begin{equation}
Y(r)=r^{\omega}\frac{1}{\beta r}\exp\left(  -\beta r\right)  .\label{yuk1}%
\end{equation}
If $I_{DPROJ}=3$, projectile density is a four-parameter Fermi function
\begin{equation}
F(r)=\left[  1+b\left(  \frac{r}{\alpha}\right)  ^{\omega}\right]  \frac
{1}{\left[  1+\exp\left(  \frac{r-c}{\alpha}\right)  \right]  }.\label{fer1}%
\end{equation}
If $I_{DPROJ}=4$, projectile density is a power of the Fermi function
($I_{DER}=0$), or its derivative ($I_{DER}=1$)%
\begin{equation}
S(r)=\frac{1}{\left[  1+\exp\left(  \frac{r-c}{\alpha}\right)  \right]
^{\omega}}\text{ ,\ \ \ \ \ or \ \ \ }\frac{dS}{dr}.\label{fer2}%
\end{equation}
If $I_{DPROJ}=5$, projectile density is a calculated from
liquid-drop model. If $I_{DPROJ}=10$, projectile density is from an
input file. \textbf{Row 12: }In case $I_{DPROJ}=1$ or 2, enter here
Gaussian or Yukawa density parameters for the projectile, according
to eqs. \ref{gas1} and \ref{yuk1}, respectively. \textbf{Row 13:} If
$I_{DPROJ}=3$ enter the four parameter of the Fermi function, eq.
\ref{fer1}. \textbf{Row 14:} If $I_{DPROJ}=4$, enter Fermi density
parameters for the projectile and its power, as in eq. \ref{fer2}.
\textbf{Row 15:} If $I_{DPROJ}=10,$ enter name (FILE3) of the input
file where the projectile density is to be found and $\alpha_{p}$ is
the proton size parameter to be used in a folding of the input
distribution with the
proton gaussian density, $\rho_{p}(r)=\left(  \pi a^{2}\right)  ^{-3/2}%
\exp\left(  -r^{2}/a^{2}\right)  $. In the input file FILE3, the
first row must give the number of the remaining rows which form an
ordered list of $r$ (in fm) $\times\ \rho_{PROJ}$ (in $fm^{-3}$) in steps of a
constant value of $\Delta r$. \ \textbf{Rows 16, 17, 18 and 19:}
Same as the previous four rows, but for the target (depending upon
the option $I_{DTARG}$).

\subsection{Eigenfunctions and energies}

\underline{Option 1} calls the subroutine \textbf{EIGEN}. The potential
parameters to build the Woods-Saxon are entered in the input file. In the
sample file above, the ground state of $^{15}$C is calculated. It is assumed
to be a 1s1/2$^{+}$ s-wave ($n=1$, $l=0$, $j=0.5$). This is just one of the
possible single particle occupancies in the ground state of $^{15}$C. The
spectroscopic factor can be adjusted to account for the occupancy probability
of this single-particle configuration.

The calculations are mainly done in the subroutine \textbf{BOUNDWAVE} which
solves the Schr\"{o}dinger equation for the bound-state problem. When
Woods-Saxon potentials are used they are constructed in the routine
\textbf{POTENTIAL}.

The output of the wavefunction will be printed in \textbf{EIGEN.TXT} and
another file specified by the user, which will be ready for use later.

\subsection{S-matrices}

\underline{Option 2} calls subroutine \textbf{SMAT}. If this option is used
the input rows for the bound-state wavefunction should be commented with a
\textquotedblleft*\textquotedblright. Here one has to specify if one wants
$S_{v}$ or $S_{c}$ and if the optical potentials are entered in a separate
file or built with the \textquotedblleft t-$\rho\rho$\textquotedblright%
\ approximation. If the later is chosen, one has to enter the options for the
nuclear densities. An option is given to include, or not, the effect of Pauli
blocking on the in-medium nucleon-nucleon cross section.

\subsection{Momentum distributions}

\underline{Option 3} calls for calculation of momentum distributions. In this
case, only the general input (charges, masses and bombarding energy) are used.
All subsequent rows may be commented, or not. One can then choose to calculate
$d\sigma/dp_{z}$, $d^{2}\sigma/dp_{t}^{2}$, $d\sigma/dp_{y}$, $d^{2}%
\sigma/dp_{t}dp_{z}$, \ or $\sigma_{elast}$.

\section{Routines included with the code distribution}

\textbf{EIGEN} finds eigenvalue and eigenfunction of a system composed of a
valence particle with charge Z1 and a nucleus with charge Z2. The nuclear
potential is a Woods-Saxon, with spin-orbit and Coulomb components.

\textbf{BOUNDWAVE} solves the radial Schrodinger equation for bound states.

\textbf{POTENT} constructs potentials to be used in the calculation of bound
state wavefunctions. Uses routine \textbf{OMP\_WS} which constructs the
Woods-Saxon potential.

\textbf{SMAT} computes the eikonal S-matrices for heavy ion collisions. The
S-matrices are functions of the impact parameter $b$. The program is
appropriate for heavy ion bombarding energies E\_lab
%TCIMACRO{\TEXTsymbol{>} }%
%BeginExpansion
$>$
%EndExpansion
30 MeV/nucleon.

\textbf{CROSS}. If IOPT = 1, this routine calculates $d\sigma/dp_{z}$. If IOPT
= 2, 3 , 4 and 5, it calculates $d^{2}\sigma/dp_{t}^{2}$, $d\sigma/dp_{y},$
$d^{2}\sigma/dp_{t}dp_{z}$, \ or $\sigma_{elast}\left(  \theta\right)  $, respectively.

\textbf{COMPDIF} is used in the calculation of the total diffraction
dissociation cross section, eq. \ref{X.4}.

\textbf{GAUST} calculates the sum over the Gaussian expansion for
$d\sigma/dp_{y}$, where $p_{y}$ is one of the components of the transverse
momentum .

\textbf{GAUSLT} is similar to GAUST, but calculates $d^{2}\sigma/dp_{t}dp_{z}$.

\textbf{FINTEG} calculates $\int_{-\infty}^{\infty}dz\ \left\vert R_{l}\left(
r\right)  \right\vert ^{2}\ P_{lm}\left(  \cos\theta\right)  $ which is used
in several other integrals. \textbf{PINTEG} deals with the integral $\int
d\phi\left\vert S\left(  \sqrt{r^{2}\sin^{2}\theta+b^{2}-2r\sin\theta
\ b\cos\phi}\right)  \right\vert ^{2}.$ \textbf{ZINTEG} computes
$\int_{-\infty}^{\infty}dz\ \exp\left[  -ik_{z}z\right]  \ R_{l}\left(
r\right)  \ P_{lm}\left(  \cos\theta\right)  $. \textbf{COMPL} is used in the
calculation of the longitudinal momentum distribution, eq. \ref{strL}.

\textbf{SMAT\_FIT} finds the best expansion parameters $\alpha_{j}$\ to fit
$S_{c}$ with an expansion in gaussians (eq. \ref{scbc}). Uses auxiliary
routines \textbf{FUNCS}, \textbf{LFIT}, \textbf{GAUSSJ} and \textbf{COVSRT}.

\textbf{FCOUL} and \textbf{CP0} are used in the calculation of the Coulomb
elastic amplitude, eq. \ref{fctheta}.

\textbf{SIGNNE} and \textbf{PHNNE} are used in the calculation of the
nucleon-nucleon scattering amplitude, eq. \ref{fnn}.

\textbf{PHNUC} calculates the eikonal nuclear phase, eq. \ref{eikphase}.

The routines \textbf{GAUSS }(eq. \ref{gas1}), \textbf{YUKAWA }(eq.
\ref{yuk1}), \textbf{FERMI }(eq. \ref{fer1}), \textbf{SAXON\ }(eq. \ref{fer2})
and \textbf{DROP} are used to calculate the nuclear densities parametrized as
Gaussians, Yukawa, Fermi, Woods-Saxon, and liquid drop functions,
respectively. The parametrizations follow the equations \ref{gas1}-\ref{fer2}.
\textbf{DROP} builds liquid drop model densities for the nuclei, following
ref. \cite{My70}.

The routines \textbf{PLGNDR, BESSJ, BESSJ0, BESSJ1, BESSI, BESSI0} and
\textbf{BESSI1} calculate Legendre polynomials and Bessel functions.

The other routines in the code are shortly described as follows. \textbf{GFV}
calculates factorials, \textbf{BETWEEN} and \textbf{FINFOUT} are interpolating
routines, \textbf{SKIPCOM} is used to skip comments in the input file,
\textbf{RSIMP} does Simpson's integrations of tabulated functions,
\textbf{FOURIER0} calculates Fourier transforms (e.g. in the calculation of
the transformed densities of eq. \ref{eikphase}), \textbf{NUCNAME} finds the
symbol associated to a given chemical element and \textbf{EXCESS} calculates
the mass excess with the parameters published in ref. \cite{AW95}.

\section{Things to do}

1 - Use the sample input file to obtain fig. 6 of ref. \cite{BH04}.

2 - Modify the input file so as to reproduce all other results published in
ref. \cite{BH04}.

3 - Use the overlap function for $\Psi_{JM}\left(  \mathbf{r}\right)  $
published in ref. \cite{NBC06} to reproduce their figure 2.

\section{Acknowledgments}

This research was supported in part by the Department of Energy under Grant
No. DE-FG02-04ER41338.

\end{document}